# Biomimetic hydrogel based on HET-s amyloïd fibers for long-term culture of primary hippocampal neurons

*Julien Hurtaud,[1] Cécile Delacour,[2] Carole Mathevon,[1] Vincent Forge[1]*

1. CEA, DSV, iRTSV, Laboratoire de Chimie et Biologie des Métaux, UMR 5249 CEA-CNRS-UJF, CEA Grenoble
2. Institut Néel, University Grenoble Alpes, CNRS, Grenoble INP, 38000 Grenoble, France

E-mail: cecile.delacour@neel.cnrs.fr

## ABSTRACT

Historically, amyloid fibers (AF) in research has always been linked to degenerative diseases. However, HET-s AF, by their morphology and function, have only little in common to pathogenic amyloid fibers such as α-synuclein or aβ and they have appeared as promising candidate for biocoating since few years. Here we have shown than HET-s amyloid fibers hydrogel is an extremely polyvalent coating material for the *in vitro* culture of primary hippocampal neurons. First, the non-cytotoxicity was demonstrated *in vitro* using standardized ISO protocols. Then, it is shown that *in vitro* culture of primary hippocampal neurons on HET-s AF hydrogels could last more than 45 days with clear signatures of spontaneous network activity, with which is a feat that not many other coatings have achieved yet. Finally, interactions between the cells, the dendrites and the hydrogels are highlighted, showing that dendrites might be able to penetrate the hydrogels in depth, therefore allowing recordings even within micrometer-thick hydrogels. In the end, those properties combined with group functionalization using standard biochemistry techniques, makes HET-s hydrogels ideal candidates to be used for the long-term growth of neurons as well as other types of cells. This versatility and easiness to use are definitely still unheard, especially for protein material. Due to its ability to transform from dry films to hydrogel when in contact with the extracellular matrix (ECM), it could also be used for *in vivo* implants, solving the issue of hydrogel damaging during the implant surgery.

## INTRODUCTION

Biocompatibility and cell adhesion are key factors for growing cell in vitro and designing electrodes for recording and stimulating electrogenic cells. The poly-lysine (PDL or PLL) coating has long been considered as the golden standard [1,2,3,4]. However, it may not be suitable option for long-term applications, as the monolayer degrades over time, potentially hindering the necessary extended investigations lasting several weeks, encompassing a broad spectrum of fundamental and applied research. To remedy this problem, other strategies have been developed to enhance cell growth and adhesion, using coating micro and nano patterned materials. Most commonly, it comprises roughening of surfaces[4,5], fabrication of pillars[6,7], grooves [8,9] or any 3D structure increasing the surface contact between the cells and the substrate[10]. Alternatively, in the last two decades, in order to avoid extensive work and cost to make micro and nanostructures from scratch, the use of materials that are by nature 3D and nanostructured gained much momentum. One of the most notorious are the assembly of carbon nanotube (CNTs), mostly used for their combination of biocompatibility[11], electron transfer mechanism, functionalization possibilities[12,13] and, of course, improve active surface to connect with cells. With all its variations, this technology has imposed itself[14], especially for contacting neurons [15]. For example, CNTs have been functionalized with carboxylic and acyl chloride groups, thus increasing the electrostatic interactions between the charged neurites and CNT surface, promoting neurite outgrowth and branching[16]. Despite many qualities, CNTs have inherent disadvantages such as high cost, toxicity issue, low-volume production and poor solubility in aqueous solvent even if the latter can be improved by mixing it with organic polymers such as poly-ethyleneimine (PEI)[17]. On another hand, organic nanowires such as polymer fibers can also be used as stand-alone for intermediate coating layer[18]. In addition, polymers such as PDMS, polyimide (PI) or polypyrrole, have been used to make pillar[19], microfibers[20], grooves[21] and even tubular scaffold[22]. If most of these polymers are insulating, conductivity can be achieved using diverse techniques such as the addition of solvents[23], thermal treatment[24] or secondary doping, like in the case of PEDOT:PSS[25]. The latter is now a gold standard for organic polymers and many applications have already been developed[26,27] such as support for the development of neuronal synaptic networks[28] or transistors for cell sensing[29]. Note that PEDOT:PSS and CNTs can also be mixed[30,31] to take advantage of both properties. However, this increases even more the complexity of the fabrication and coating, lowering the transferability of such technology to common use. If they dispose of many advantages, these polymers also have their own limitations, often displaying low stability in water, complex fabrication and high cost. The biocompatibility of PEDOT:PSS is also still discussed among the community as it is known to release some toxic compounds used during its fabrication[32], possibly interfering with the objects of the study and the devices.

As a way to avoid any issues with biocompatibility, a promising approach relies on the use of natural proteins such as fibroin[33,34] or collagen[35,36]. More than simply promoting neuronal network development, they are able to successfully support 3D culture of primary neurons, necessary to make in-depth connectivity study for memory and learning behaviour[37]. The structuration of these natural fibrillar materials can also be used to guide and promote neurite growth[38]. Not costly and efficient in their patterning abilities, they are however insulating, thus limiting their use as coating materials with electrodes. We propose here the use of amyloid fibrils for the surface coating. They have several potential advantages. As they are the results of protein self-assembling, they allow the design of interfaces by bottom-up approaches. They have intrinsic long distance ionic/protonic conductivities and they can be functionalized to have also long-range electronic conductivity[1]. Various functionalization's can also be envisaged for a better control and/or to obtain specificities of their interactions with cells. As amyloid fibrils were first described in the context of neurodegenerative diseases[39,40], their biocompatibility could be questioned. However, several functional amyloid fibers have been described in all kinds of organisms[41–43]. The amyloid fibrils used for this work are made of the prion domain of HET-s. These are functional amyloids involved in the self/non-self-recognition of the filamentous fungus *podospora anserina* [44]. The prion-domain of HET-s, made of the $219^{th}$ to the $289^{th}$ amino-acid (figure 1a), self-assembles into well-defined nano-fibrils at pH 4 (figure 1b) [45]. The high-resolution structure of these nano fibrils has been resolved by solid-state Nuclear Magnetic Resonance[46], allowing the rational engineering of these fibrils. Surface coating can be easily performed by drop casting these nanofibrils, letting them dry and, then, by dipping them within a solution at higher pH (such as a culture medium). Then, a stable and adhesive hydrogel made of intricate and laterally assembles nanofibrils is formed at the surface[1,47,48]. The use of this hydrogel has been explored as a nanostructured coating for the long-term *in vitro* culture of primary hippocampal neurons. The hydrogel coating derived from HET-s amyloid fibers have been deposited in several conditions on conventional growth substrates and on microelectrode arrays to monitor both the outgrowth and the electrical maturation of primary mammalian neurons *in vitro*. Confocal and scanning electronic microscopies were additionally used to assess further the interaction between the intricate mesh of neurites and the amyloid fibers. The hippocampal neurons cultured on HET-s hydrogel followed the several maturations steps up to the establishment of electrical activity, which could last 45 days up to the end of the culture period, confirming the potential of HET-s hydrogel coating for the long-term maintenance of cell adhesion and electrical communication. Due to its ability to transform from dry films to hydrogel when in contact with aqueous solution of pH 7, which is the case of extracellular matrix (ECM), it could also be used for *in vivo* implants, preventing tearing effects during implantation.

# RESULTS AND DISCUSSION

**Manufacturing nanofibril hydrogel coating from HET-s.** The HET-s prion-domain was produced as recombinant protein and the self-assembling into nanofibrils was induced by an incubation at pH 4 (materials and methods)[1,47]. The biocompatibility of amyloid fibers made of HET-s prion-domain was first assessed using an immortal fibroblast cell line (L-929) as required by the ISO 10993-5:2009 "Biological evaluation of medical devices" about cytotoxicity *in vitro* tests for medical devices. HET-s AF were dosed at different percentage in Eagle Minimum Essential Medium (EMEM10). Then mammalian cells were cultured with this reagent. Positive (polyurethane film containing 0.1% zinc diethyldithicarbamate) and negative control (high density polyethylene sheet) were prepared with the same protocol by replacing amyloid fibers with their respective materials. The cell viability was then calculated through optical density comparison (materials and methods). Regardless the percentage of HET-s AF hydrogel, L-929 mammalian cells had 100% viability and no cytotoxic potential was found at all (Table 1), confirming the biocompatibility of HET-s AF and their suitability to be in contact with living tissues.

| Material | Percent Viability of Control Articles | System Suitability |
|---|---|---|
| Positive control | 3 % | Met criteria |
| Negative control | 97 % | Met criteria |

| Material (Dilution HET-s) | Percent Viability of HET-s AF | Cytotoxic Potential |
|---|---|---|
| 100 % | 103 % | No cytoxic potential |
| 50 % | 100% | No cytoxic potential |
| 10 % | 100% | No cytoxic potential |
| 1 % | 97 % | No cytoxic potential |

**Table 1:** Viability test of HET-s AF amyloid fibers on L-929 immortalized mammalian cells (type fibroblasts) following the ISO 10993-5:2009 standard for the "Biological evaluation of medical devices". For all conditions, cell viability was close to 100% with therefore no cytotoxic potential.

**Structural properties of HET-s AF hydrogel.** Transmission electron micrograph (figure 1b) shows the typical shape of the nanofibrils, with a mean diameter and length of around 5 nm and 10 μm. Isolated amyloid fibers are clearly distinguished, with no sign of aggregation or amyloid plaques as with other pathogenic AF. It is worthy to notice the exceptionally high aspect ratio of the obtained nanowires, reaching 2000. For comparison, most of multiwalled carbon nanotubes (MWCNTs) commercially available exhibit an aspect ratio lower than 1000. Other bio-nanomaterials such as actin and microtubules also have aspect ratio lower than 100. Homogeneous coatings of HET-s AF were successfully obtained and characterized as function of the HET-s concentration the expected amount of amyloid fiber in the mixture (figure 2). To this purpose, the HET-s AFs were stained with Thioflavin-T (ThT) protein which fluorescence in the green region, is greatly enhanced when bound to amyloid structures. A concentration of 50 µM in ThT was set in the initial dispersions of HET-s AF at 300 µM (1:1), 100 µM (1:3), 30 µM (1:10) and 10 µM (1:30). After drop-casting, drop was left to dry in ambient environment (23 °C / 40 % RH). Once in thin films, it swelled by putting it in direct contact with PBS 10 mM (pH = 7,4) and their fluorescence were studied using confocal microscopy. While confocal resolution (few hundreds of nm) is not sufficient to observe individual AF (10 nm in diameter), it enables to observe protein coating meso-structuration. The fluorescence intensity was normalized to focus on the distribution of the materials. Thus, the first observation, not visible on the figure above, was a steady decrease of fluorescence intensity (10 times lower from 1:1 to 1:10), logically coinciding with the decrease of HET-s material as concentration decreases. Concerning the HET-s hydrogel thickness itself, surprisingly, for all conditions, it was comprised between 15 and 30 µm. More interestingly, the confocal images have shown major differences in the density and homogeneity of the coating which clearly depends on the HET-s concentration. A transition from a continuous film to isolated aggregates have been clearly observed when reducing the concentration of HET (from left to right figure 2). At 1:1 concentration (300 µM), HET-s hydrogel was homogeneously spread across the glass slide. For 1:3 and 1:10 dispersions, it is much more heterogeneous with the apparitions of what seems to be meso-filaments of proteins, with a succession of protein dense and empty zones. Finally, for 1:30 conditions, image showed only few AF left highlighting a discontinuous and very partial HET-s coverage on the surface.

**Electrical properties of HET-s coating.** Because of its use as an intermediate layer to allow and to promote the growth of primary neurons on electrical neural interface, it was mandatory to make sure that the HET-s AF coating does not act as an insulating layer between the cell and the recording device. Electrochemical Impedance Spectrometry (EIS) technique was used as a way of probing the charge transport properties of the HET-s AF film on a wide range of frequencies, from 10 Hz to 400 kHz (Supp. Info. 1). Experiment was made on the same setup as in the next experiments, i.e. HET-s hydrogels were prepared directly onto commercial MEA

(MultiChannel Systems – 60MEA200/30iR-Ti) and PBS was added into the well to connect the counter electrode.

As a result, the impedance measurement showed that adding HET-s AF hydrogel does not increase the impedance of the microelectrodes, but on the contrary reduces it by a factor 2 from 10 Hz to 100 kHz. This result demonstrated first the conductive nature of the HET-s AF hydrogel and its ability to improve the impedance of the microelectrode. This enhancement is surely due to the intrinsic charge transfer of HET-s AF relying on protons and ions, therefore supporting long range transmissions of charges from the neuron vicinity to the electrode. This property, very rare for protein-based materials makes HET-s hydrogels a really unique material, opening new perspectives for hydrogel coatings materials. The graph on the right of figure S1 also shows the dispersion of values for more than a twenty different HET-s coating. Once can observe that the values are very consistent for all frequencies.

**Primary hippocampal neuron cultures** have been assessed on HET-s AF hydrogel to evaluated its potential as a growth substrate for neural interface. To this purpose, hippocampal primary neurons extracted from mouse embryo have been seeded on the coated sample as function of the amyloid fibers concentration. The impact of HET-s AF on the cell development was assessed by following both structural and functional changes *in vitro*. To that end, the electrical activity and morphology of primary hippocampal neurons were monitored during the three weeks which is the typical maturation time. These features have been compared with controls, obtained from poly-L-lysin (PLL) coated samples which is an adhesive protein widely used for *in vitro* neuronal culture. For the two batches of HET-s AF and PLL-coated samples, the morphology of neurons was observed on glass coverslips while their electrical activity has been recorded with array of microelectrodes. In particular, a drop of HET-s AF suspension was casted onto glass coverslips and six arrays of 30 μm-wide TiN multielectrode (MEA from MultiChannel System 60MEA200/30iR-Ti) with a mother solution at 300 μM. Four conditions were made: at a concentration of 300 μM (1:1), 100 μM (1:3), 30 μM (1:10) and 10μM (1:30) in hydrochloric acid (HCl) pH = 4-5. Once the drop dried in sterile condition under the biosafety cabinet, neuronal culture medium was added (pH = 7.4) to transform the HET-s AF dry film into its hydrogel form. Indeed, in a neutral pH aqueous medium, it swells into a three-dimensional fibril network, stabilized by hydrogen bonds and electrostatic interactions. These hydrogels also have the property of sticking to the surface it is attached with no diffusion into the external medium. Finally, dissociated primary hippocampal neurons extracted from mouse embryo were seeded onto the HET-coated glass coverslips and MEA's well. To follow the adhesion and neurites outgrowth along the culture time, glass samples were successively fixed in paraformaldehyde (4%) and then neurons were immunostained with anti-alpha tubulin (1:1000 sigma aldricht), phaloïdin (1:300 Sigma Aldrich) and DAPI (1:1000, Millipore) to label the cytoskeleton and the soma respectively (figure 3a-d). Anti-camK2 and anti-GAB primary antibodies were used at later maturation stage to discriminate excitatory and inhibitory neurons (figure 3e-f).

As illustrated within figure 3 and figure S5, neurons efficiently attached on the HET-s films, while bare glass does not promote the neurons adhesion as shown at the edges of the HET-s coating (indicated by the dashed line). Also, for the 1:30 HET-s dilution, cells did not attach well, clustered and died after few days (figure S6), therefore this condition is excluded from the following results. On HET-s AF substrate from 300 µM to 30 µM, the number of attached neurons is about 1100 cell/mm$^2$ which gives about 230 000 cells for the 200 mm$^2$ glass slide. It agrees with a seeding density of 500 000 cell/mL for 500 µL deposited, giving a total of 250 000 seeded neuronal cells. Thus, most of seeded neurons have attached, demonstrating the highly efficient adherence properties of the HET-s AF material. Around 2-4 neurites per soma are observed after two days *in vitro* (average 2.8 neurites per neuron), which is as expected for hippocampal neurons. Together, these results demonstrate that HET coating promotes neurons adhesion and neurite spreading at the early days of the culture. Later (from DIV 5 to DIV 30), the reconstructed networks of neurites continue to spread, covering the overall surface (80% at DIV5, figure 3 and figure S5), which demonstrate the high neuronal affinity of HET-s AF. The absence of neurites' bundle and somas' cluster, for 300 µM, 100 µM and 30 µM solutions, confirm further the healthy state of neurons grown on HET-s. The neurons were kept in culture several weeks ( > 45 days ) until we the end of the culture, with no signs of degeneration. From DIV8 to DIV21 we followed the expression of two enzymes (CaMKII and GAD67) expected to be expressed in excitatory and inhibitory hippocampal neurons respectively. Both are important in the establishment of neuronal activity and long-term potentiation. At DIV8 , GAD-positive staining was found in most of the neurites, and the ratio of soma positive to GAD over CamKII is about $N_{GAD}/N_{CamKII} = 0.68$ after 21 days. The expression of CaMKII within the neurites increased with the culture time as observed on the micrographs obtained at DIV21 (figure 3e-f). This confirms further the electrical maturation of the reconstructed network of neurons, which is as expected for primary neurons maintained 3-weeks in culture. Overall this series of experiment suggested that HET-s AF, at concentration equal or above 10 µM, are efficient candidate as a coating for neuronal cultures in systems such as glass coverslip and MEAs.

**Electrical maturation on HET-s AF coating.** In order to further investigate the impact of the coating on the cell development, quantitative measurement of the neuronal activity was performed with the microelectrode arrays (MEAs) that enabled multisite recording of neural spike along the culture time. With the same protocol as in the previous experiments, the samples were coated with HET-s AF at different concentrations (300 µM, 100 µM, 30 µM and 10 µM) which corresponds to a HET-s thickness of approximately 30 µm, (figure 2). Neurons were seeded at a density of 250 000 cell/mL. Electrical recordings were performed in the culture medium at room temperature, under a biosafety cabinet. Recording conditions were keep the same for each recording session. The acquisition was started 5 minutes after placing the MEA in the recording setup to let the cells recover, and last for 60 seconds.

According to the table figure 4a, the onset of neuronal activity appeared seven days after seeding the cells. Then the number of active electrodes and spiking rate both increased with the culture time. After two weeks, the samples coated with a HET-s concentration of 100µM and 30 µM (1:3 and 1:10 respectively) exhibited neuronal activity which is coherent with the expected maturation time of neurons. At DIV15, most of *in vitro* neural network is expected to be electrically active. On the other hand, no activity was measured on the samples at 300 µM and 10 µM (1:1 and 1:30 respectively). Regarding the active MEAs, after 14 days, the number of active neurons and the spike rate appear to depend mostly on the coating and the sensors efficiency. For the 1:3 HET-s dilution, the detection rate decreases at DIV19 and thus only the samples coated at 1:10 HET-s dilution (30 µM) still detected neural spikes. Remarkably, out of the three cultures plated with 1:10 diluted HET-s hydrogel, two of them showed active electrodes after DIV29 and one of them even lasted more than DIV35, enabling long-term *in vitro* culture. Figure S3 shows the shapes of the recorded spikes, both amplitude and duration are as expected for hippocampal neurons spontaneous activity. As said previously at 1:30 dilution (10 µM), neurons have barely attached and developed, forming clusters and neurites bundles, thus showing degeneration signs. At DIV14 and above, most neurons did not survive and no electrical activity was detected (figure S5). On the opposite, the 1:1 and 1:3 dilutions showed no sign of degeneracies or toxicity (e.g. bundle and soma cluster or membrane disruption) on the immuno-staining micrographs: the dendritic tree was dense and widely spread over the microelectrode's arrays (figure S5). Thus, neurons networks were comparable in all respects to the one grown on the thinner HET-films, at 1:10 dilution. Thus, the non-activeness on denser coatings seems rather stem from the efficiency of the cell-device coupling rather than from the electrical state of neurons.

Typical shape of action potential are shown figure 4c (1:10 dilution, 30 µM). The time trace of the extracellular voltage exhibited the highest spike amplitude, being about 100 µV peak-to-peak ($V_{pp}$) at DIV14 and 300 µ$V_{pp}$ at DIV29. Also, these MEAs provide the highest time-stable operation regime, enabling to record neuronal activity still after 35 days in culture (figure 4b). The last recordings display detectable activity at DIV48. To our knowledge, this is barely achieved with conventional protein-based hydrogels and very rarely with inorganic materials. Overall, it has also shown a better stability that PLL coated substrate (typ. DIV15-21), which is the standard material used for such *in vitro* neuronal cultures. PLL control timestamps and recorded signals are available figure S3-S4.

Given the differences of neuronal viability and action potential detectability, this assay highlights an optimal concentration close to 30 µM of the HET-s hydrogel for the long-term growth and the detection of neural spikes.

Spike detection is greatly reduced for concentration about of 100 µM and more, even if the neuronal display healthy morphology. The most probable explanation is linked to the thickness of the HET-s AF hydrogels acting as a physical barrier between the target neuron and the sensing electrodes up to a certain extent. Additionally, as HET-s AF are good protonic and ionic conductor, as seen previously with electrochemical impedance spectroscopy (EIS) experiment, using it as a whole layer might short circuit MEA electrodes, redistributing the charges from the neurons in the whole layer AF. From this observation, the topological and physical interaction between the HET-s hydrogel and neurons should be further assessed in a future sutdy.

**Topology of HET-s AF hydrogel and neuronal culture** was assessed by scanning electron microscopy (SEM) which enabled to distinguish the nanomesh of HET-s amyloid fibers on which neurons were growing. The drawback of this technique is however the sample preparation that requires dry samples. Therefore, the micrograph should only be interpreted as vertical projection of the 3D culture and hydrogel. Precise preparation protocol can be found in materials and methods. The micrographs show neurons grown on HET-s 1:10 (30µM) hydrogel and on PLL coated samples for comparison. Samples were fixed in time at DIV4, DIV6 and DIV12, respectively representing early development stage and mature neurons.

As illustrated in figure 5, the HET-s AF coating could be clearly distinguished as well as the neuronal bodies. On the figure 5a, c, and e the SEM micrographs show hippocampal neurons cultured on the 1:10 HET-s AF hydrogel and the figure 5b and d are obtained on standard PLL monolayer (neurons from same culture). With HET-s hydrogel coating, even if the mat is extremely dense (30µM), individual nanowires can easily be discerned (figure 5a). Their lengths vary approximately between 300 nm and 1,5 µm which is less than expected values[48]

(about 10 μm) certainly because the micrographs only reveal the emerging part of HET-s AF. These images confirm that, at least until DIV 6, neurons are located above the substrate, thus using the nanostructured substrate as a base for development. At this stage, neurons look healthy, displaying nominal cell body morphology and dendrite/axon development.

Figure 5a highlights a lamellipodium extension over the HET-s AF hydrogel showing that this neuron was well attached and still exploring its environment at this stage. Comparing it to the PLL controls, the absence of differences until DIV4 demonstrate that neurons are indeed well developed and healthy in early stage. At DIV 12, compared to the control made on PLL (figure 5d), we see that some neurons soma and most dendrites have entered the HET-s AF layer. Indeed, the intricate dendrite network is much less visible that PLL control and round traces, the size of neuron's cell body, can be distinguished at the surface of the film (figure 5e). We also observe film fractures; we attribute these to SEM sample preparation process, requiring freeze-drying step and degrading the whole sample which is fragile to such processes. It shows that HET-s AF substrate is not only a nanostructured mat able to support neuronal growth, the hydrogel moreover acts as a mattress possibly allowing the formation of a 3D culture, making it is the only natural material featuring this property so far in the literature.

Confocal fluorescent imaging was finally performed on the HET-s/neuronal culture at DIV13 to observe the same samples but in hydrated condition within liquid solution. Immunostaining and imaging protocols are detailed in materials and methods. Briefly, hippocampal primary neurons were cultured on a hydrogel deposited from a 30 μM (1:10) dispersion of HET-s nanowires on a glass coverslip (same deposition condition and culturing protocols than previously). Using the Z-stack mapping of the confocal microscope, the samples were observed in the z-direction at several hydrogel depth value. The reconstructed 3D image of the probed section shows scans taken at different z-position of the three-dimensional hydrogel/cell sample, providing an image over the whole sample thickness (figure 6). It should be noted that the fixation technique might have squeezed the culture between two glass slides; therefore, the height values may not be accurate.

Within figure 6a , the substrate interface or ground level (Ref - 0μm) consists only of dendrites and axons while the cell bodies of the neurons are observed a few micrometres above. On the side slides of the second culture (figure 6b), there are clearly neurons at different heights in the culture. Unfortunately, we were not able to image the AF in the same experiments as both fluorescences are emitting in the same range and the dye fluorescence has much higher intensity than AF intrinsic fluorescence, so it is not possible to clearly co-locate the neurons and the hydrogel AF matrix. However, as we know that neurons cannot grow on bare glass, there are only two possibilities: i) either the hydrogel is under the culture or ii) the culture is housed within the hydrogel. The fact that the cell bodies of the neurons are not all on the same plane supports the latter hypothesis. It is also possible that the surface of the HET-coating is not flat, as shown in figure 1d. Our hypothesis still needs to be confirmed as the thickness of the hydrogel is of the same order as the diameter of the soma of a neuron, which makes it difficult to draw a definite conclusion about the entanglement between the neurites and the HET nanowires and the 3D nature of the culture. If it is still impossible to quantify this interaction, we can nonetheless say that there is a deep interaction between the two, which could be expected. The extent of this interaction is still be determined, if it is only mechanical or other phenomena happen such as electrical/ionic interaction.

**Preliminary study in vivo.** All experiments related to this article were made *in vitro* and as said earlier, the purpose of this technology is to be transferred for neuroprosthetic implants. For the next series of experiments, it was therefore fundamental to study the biocompatibility *in vivo*, that is much harder to achieve as living tissues comprise a wider diversity of molecules, enzymes and cell types such as astrocytes, fibroblasts, and because the implantation (surgery and the implanted devices) induce damages that are difficult to distinguish from that of the implant coating.

The campaign was performed using standard rules detailed on a cohort of rats. The test consisted in implanting a cylindrical-shaped mandrel coated, or not in the case of control population, with HET-s AF at a concentration of approximately 300 µM. Three parameters were evaluated during the campaign: weight, behavior and physical appearance. For each parameter, the animal was given a clinical score every week from 0 to 3 which corresponded to specific changes (figure S2). The study lasted for two months, comprising the month of accommodation. After one month of implantation, rats were euthanized for the histology to be made. It comprised staining of brain sections with Cresyl Violet, glial fibrillar acidic protein (GFAP) and Ionized calcium-binding adaptor molecule 1 (Iba1). Mandrels with and without HET-AF hydrogels were studied. Characteristics of the mandrel (dimensions, shape and material) make it a negative control, inducing almost no inflammatory reaction.

After the implantation, all scores remained at 0, showing no change in whether weight, behavior or physical appearance. As for histology, each marker reveals its own piece of information. The first is cresyl violet, which reveals cellular architecture of nervous tissues, therefore marker of cell proliferation and necrosis around the implantation site. The result of this experiment showed that for all cerebral hemispheres implanted with HET-s AF coated mandrels, the amount of proliferation or cell loss was similar (marked with the letter M and delimited by a dotted line on figure 7d) to that of the control bare stainless-steel mandrels (figure 7c). Looking at the images closely, it seems that there is even less cell proliferation at the interface with the mandrel coated with HET-s. Overall, these results suggest that amyloid fibers appear to be, at least, very well tolerated by the cerebral parenchyma for a contact duration of one months.

Other than cresyl, two immuostraining were made to characterize inflammatory response after the implantation. The first targeted the glial fibrillar acidic protein (GFAP) which is an intermediate filament present in glial cells of the central nervous system, and in particular astrocytes. Astrocytes are key components of gliosis, healing process after brain injury. The immunostaining directed against the GFAP protein showed that for all the cerebral hemispheres implanted with the control stainless steel mandrels, only little proliferation or astrocyte morphological modification was observed around the implanted area (marked with the letter M and delimited by a dotted line in figure 7e). For the cerebral hemispheres having received the stainless-steel mandrels coated with amyloid fibers, the results obtained were comparable to those of the control condition (figure 7f). This result suggests that the HET-s AF hydrogel do not induce an inflammatory reaction after a period of contact with the brain parenchyma for 1 month. The second targeted Ionized calcium-binding adapter molecule 1 (Iba1) protein. Among brain cells, the Iba1 protein is specifically expressed in microglia. Microglial activation is another type of inflammatory reaction that involves the activation of microglia, the immune cells in the brain. Similarly, to GFAP test, for the cerebral hemispheres having received the stainless-steel mandrels coated with HET-s AF, the results obtained were comparable to those of the control condition (figure 7g-h).

As a result, the implantation of the mandrel coated with amyloid fibers indicate no sign of any tissue damage nor inflammatory reaction specific to HET-s AF contact after 1 month of contact with the brain parenchyma. In fact, it seems that the hydrogel may have reduced these inflammatory and immune reaction. Although this should to be confirmed with further statistical studies but which are beyond the scope of the present study. These explorations have demonstrated an adapted response of the host tissue and demonstrate the implantable nature of amyloid fibers at a cerebral site and allow their use in implantable devices to be considered.

**DISCUSSION**

By the combination of all elements mentioned throughout this article, one can grasp the potential of using protein nanowires and here especially HET-s AF as an intermediary material to interface electrodes or inert surfaces and biological tissues. With non-cytotoxicity proved *in vitro* as well as its ability to transport ionic charges of from the cells to the electrodes, HET-s hydrogels have proven to be extremely efficient for the growth and development of healthy neuronal and cellular cultures. It goes even further than cultures coated with PLL by detecting action potentials up until 45 days in culture with healthy electrical and structural characteristics and regular signal to noise ratio.

SEM and confocal microscopy enabled further characterizations of the intricate network of neurites and amyloid nanofibers. This ability to resolve both proteins fibers and neurons on the same images puts in evidence that neurites and soma could penetrate the hydrogel to a certain extent in the z-stack reconstitution. It poses the question on the origin of the electrical signals detected at the electrodes: do axons and dendrites reach physically the electrode? Is the HET-s hydrogels able to transport charges? Most certainly, the recorded electrical signals are combination of both phenomena but deciphering the percentage for each contribution is extremely difficult. Still, for distances over to 100 nm, the field-effect capacitance detection principle is becoming negligible, and chances are low that axons to completely pass through the whole hydrogel thickness. This is why the hypothesis of a predominant detection assisted by ionic/protonic properties of HET-s AF is privileged. Impendence measurements (figure S2) also confirmed the contribution of HET-s AF to reduce the microelectrode impedance at the liquid interface.

Aside the novelty of using only proteins materials as growth coating and charge transporter on primary neurons, amyloid fibers have others advantages for applications in brain-machine interfaces that have not been yet implemented and thus represent a future niche for improvement. On a materials' point of view, its ability to turn from dry thin film into swollen hydrogel once in contact with neutral (pH = 7) aqueous media is extremely useful if used as coating on an implant. Indeed, as brain fluids are typically in that required range, it solves the problems related to hydrogels getting peeled off during the implantation because of frictions against the wound outer surface. With HET-s coating, the hydrogel would form only once settled at its final location, therefore adapting and fitting perfectly the shape of the injury.

As proteic material, we could also easily imagine the patterning of the hydrogel using UV light and mask to remove hydrogels at desired locations such around the sensing sites to increase the sensing device performances.

Another aspect lies in HET-s AF surface that can also be functionalized easily. Its auto-assembly process allows an even easier functionalization as the process is made on monomeric protein in solution which will later turn into fibers and hydrogel, such as RdHET-s, a chimera protein fusing HET-s proteins and Rubredoxin for long range electron transfer between bounded enzymes and electrode's surface[1]. One could imagine transposing this example for functionalization: increase adhesion through the binding of cadherin proteins to fibers, increase conductivity of the hydrogels using other types of metalloproteins like cytochromes-c.

**CONCLUSIONS**

The reported studies have shown that not only HET-s amyloid fibers based materials are biocompatible but are also able to act as effective coating for primary neuronal growth *in vitro*. Indeed, the cultured neurons on HET-s hydrogels are developing even in a healthier way than controls, and the detection efficiency is enhanced as well, enabling to record physiological action potentials up to 45 days. Moreover, it seems that the 3D structure of the HET-s AF hydrogel allows neurons soma and dendritic extensions to settle in, furthering and improving the anchorage of cells in the material. Overall, we observed that HET-s AF hydrogels combine all properties necessary to be implemented in hybrid devices that takes advantages of both organic biomaterials and microelectronics for highly efficient cell/neuron culturing and sensing: biocompatibility, ionic/protonic conductivity, mechanical properties similar to biological tissues and endless functionalization possibilities. These results are a first step toward innovative and all-in-one materials and one can only see further development in this field, which is necessary to offer alternatives to classical coatings methods suffering from limited property coverage, superposition of materials layers and complexifying the whole system.

**Figure 1**: A) Amino acid sequence of HET-s(218-289) prion domain protein. B) 3D visualisation of HET-s structural folding. Every protein is composed of 5 vertically stack sub-units, made of two successive β-sheets. B) C) TEM images of HET-s amyloid fibers. Scale bar 200 nm (top), 50 nm (bottom). This protein does not form any aggregates and form nanowires with the following characteristics: 10 µm in length, 5-10 nm in diameter.

**Figure 2:** Confocal microscopy images of HET-s hydrogels at several concentrations (300 µM, 100 µM, 30 µM, 10 µM respectively from left to right). The coatings are stained with the AF-specific ThT dye ($\lambda_{exc}$= 458 nm - $\lambda_{em}$= 460 nm to 550 nm). Scale bar 100 µm. Under each image a corresponding scheme represent the thought arrangement of AF within the whole layer of the hydrogel. Lower concentration induce heterogeneity in the layer. The remaining AFs assemble into macrobundles, leaving zones with much lower density. At 10 µM concentration, it seems that fibers do not form proper macrobundles, resulting in a very scarce coating.

**Figure 3**: A-D) Epifluorescence micrographs of primary hippocampal neurons seeded on glass coverslips partially covered with HET-s AF hydrogel at DIV 1 (A-B) and DIV 2 (C-D). Neuron markers YL1/2-FITC (green) labels the neuron cytoskeleton, DAPI (blue) the neuron's soma and phalloidin TRITC (red) the growth cone and residual astrocytes. Neurons exhibit a healthy early development on all HET-s AF substrates. Scale bar 10 µm. E-F) Epifluorescence micrographs made on primary hippocampal neurons seeded onto MEAs at DIV21: (E) 300 µM and (F) 30 µM HET-s AF hydrogel. Neuron markers: YL1/2-TRITC (red) marks neurites, GAD (blue) the inhibitory neurons and CamK2 (green) the excitatory neurons. For both conditions, the culture display a healthy ratio between inhibitory and excitatory neurons. Moreover, neurons show good connectivity, distribution and morphology. Scale bar 5 µm.

**Figure 4**: A) Table representing the number actives electrodes onto 64 electrodes MEA plated with primary hippocampal neurons and then coated with HET-s AF of different concentrations. Mother solution (1:1) has a HET-s AF concentration of 300 µM. Thus 1:3 = 100 µM / 1:10 = 30 µM / 1:30 = 10 µM. This table shows for that unitary spikes recording, the 1:10 dilution seem to be optimal. B) Superposition of spikes recorded on an electrode of a MEA coated with HET-s AF at DIV 35. Primary component analysis enables to distinguish four shapes, corresponding to the spiking events from four individual neurons. Amplitude and shape are as expected for neural spikes. The signal- to-noise ratio is also very high with the highest values peak-to-peak reaching approximately 400 µV. Having such signal on such mature neurons highlights the potential of HET-s coating for long-term in-vitro culture. C) Examples of recorded MEA signals (2 for each DIV) at DIV 7, DIV 14, DIV 29 and DIV 48. All were recorded with MEA coated with 1:10 HET-s AF dilution, ,i.e.30 µM. At DIV 7, neurons are not connected yet and thus no signals are recorded. From DIV 14, many spikes are occurring with active and passive phases. At DIV 46, thus extremely mature culture, spikes of lower amplitude can still be easily observed. Zoom on spikes are available in Supp. Info (S2) and CTRL PLL traces are also available within Figure S4.

**Figure 5**: A) SEM image of primary hippocampal neurons seeded on 30 µM HET-s AF coating. Neurons' body can be seen on the bottom left corner and HET-s AF are distinguishable on the background substrate. Image made at DIV 6, i.e. early stages of development. Scale bar 500 nm. B) and C) SEM images of the same type of neurons on Poly-L-Lysine (PLL) (B) and HET-s AF (C) at DIV 4. Scale bar 10 µm. D) and E) SEM images of the same type of neurons on

Poly-L-Lysine (PLL) (D) and HET-s AF (E) at DIV 12. Scale bar 10 μm. Topology of the surface is very different between both conditions. In the case of HET-s coating, it seems that some neurons are embedded under the fractured top surface (surface of dried HET-s hydrogel). If the SEM preparation process completely dry the sample (provoking fractures as seen in (E)), this is still a strong hint that from certain early DIV (>6), neurons might penetrate HET-s hydrogel, suggesting a strong interactions between them and thus explaining why impedance is lowered by the addition of HET-s AF.

**Figure 6**: Images of rats brain implanted with control (A) or HET-s AF (B) coated mandrels M in the brain region of parenchyma. Brains slices at the location of the implant, coated with HET-s (right) or not (left), were marked with cresyl violet (C,D), GFAP (E,F) and Iba1 (E,F). Cresyl violet colors cell endoplasmic reticulum, useful to observe cellular architecture, cell proliferation and cell necrosis. On the other hand, anti-GFAP and anti-Iba1 immunochemistry respectively mark glial cells such as astrocytes and microglial cell which are activated during inflammation processes. Overall, this in vivo biocompatibility test shows no additive cell necrosis, immune nor inflammatory response specific to HET-s AF coating of inserted mandrel.

**Figure 7**: A) Confocal microscopy images of hippocampal neuron culture on HET-s substrate at DIV13. Cells are labelled with TRITC (Exc 495 nm / Em 519 nm) marking cell's cytoplasm. Successive images are extracted from Z-stack made on the culture. Height increase from left to right (see top left corner). One can see that the culture is composed of a mat of neurites at the "floor" z-level and neurons' soma few micrometers up. B) 3D reconstruction from x40 confocal Z-stack of a similar culture as A). Left: xz plane is swiped on the y-axis. Right: yz plane is swiped on the x-axis. Neurons are indeed located at different layers C) z projection of the 3D reconstitution of the same image set as B). Scale bar 5 μm.

**Materials and methods**

*Production and purification of HET-s amyloid fibers.* HET-s(218-289) protein was expressed in E. coli BL21(DE3). After expression, bacterial pellets were resuspended and sonicated in 50 mM Tris/HCl pH 8.0, 150 mM NaCl and 6 M guanidine hydrochloride. The supernatant was cleared by centrifugation at 40,000 g for 1 hour. The protein was purified from the supernatant by affinity chromatography on Ni Sepharose column (GE Healthcare). The buffer was exchanged by dialysis from denaturing to renaturing condition with Spectra/Por7 dialysis membrane MWCO 1 kDa. The solution used for dialysis were Tris 10mM pH 7.5 or HCl 0.1 mM pH 4. After dialysis, protein sample concentration was modified by dilution or by concentration on Amicon Ultra Centrifugal filters (Merck Millipore). At acid pH and high concentration (300 μM) a gel of HET-s amyloid fibers was obtained.

*HET-s deposition and cell culture.* Glass coverslips and microelectrodes arrays were cleaned in acetone, ethanol, isopropanol and dry in nitrogen. Surface were activated by an oxygen plasma and finally exposed 100 μl or 500 μl of HET-s AF (1mM diluted in KCl (pH=*) or poly-L-lysin for control samples (0.1 mg/ml final concentration). Primary neurons were extracted from E16 mouse embryos hippocampi, dissociated mechanically after incubation of 10 min in trypsin solution as described in our reported protocol. (Veliev Biomaterials 2016) Cells were counted and diluted to 500 000 cells/ml in MEM solution supplemented with 10% horse serum, 1% glutamine and 0.5% peniciline/streptomycine (all cell culture medium and supplements were supplied by ThermoFisher Scientific Inc.). 500µl and 1ml of cell suspension were seeded over the glass coverslip and microelectrode arrays respectively. The samples were transferred in a humidified 5% $CO_2$ incubator for 2-3h, then the attachment media was replaced by glia-conditioned neurobasal media for maintenance of cells. Media was partially changed once a week by removing 1/3 and adding 2/3 of the solution to compensate evaporation.

*Fixation and labelling (samples on glass slides).* Cells were fixed with 4% paraformaldehyde solution for 15min, then rinsed three times with phosphate buffer solution. Immediately after, neurons were permeabilized in PBS-0.25% Triton X100, blocked in PBS-2% bovine serum albumin and immunostained with anti-alpha tubulin (1:1000, Millipore), anti-phaloïdine (1:300 Millipore) and DAPI (1:1000 Millipore), anti-CanK2 (1:300 Millipore) and anti-GAD (1: 300 Millipore) to label the cytoskeleton, actin filament, the soma, excitatory and inhibitory neurons respectively. Incubation of primary anti-bodies last for 2h at room temperature. Then samples were washed and incubated with the secondary anti-bodies for 2h, in dark at room temperature. Samples were rinsed two time in PBS and lastly in water, and then were mounted on glass slides and imaged using an Andor Zyla scientific camera, controlled by Andor Solis and an Olympus BX51 microscope. The images were processed with ImageJ.

**Supporting Information**
Supporting Information is available from the Wiley Online Library or from the author.

**Acknowledgements**
Authors thank Gael Moireau and Pierre Gasner for their assistance in the management of cell culture room BIOFAB where the experiments were realized. Authors acknowledge grants from the French National Agency of scientific Research under the projects ANR-18-CE42-0003 NANOMESH

**A**

```
     217       227       237       247       257       267       277       287
MKIDAIVGRNSAKDIRTEERARVQLGNVVTAAALHGGIRISDQTTNSVETVVGKGESRVLIGNEYGGKGFWDNHHHHHH
```

**B**

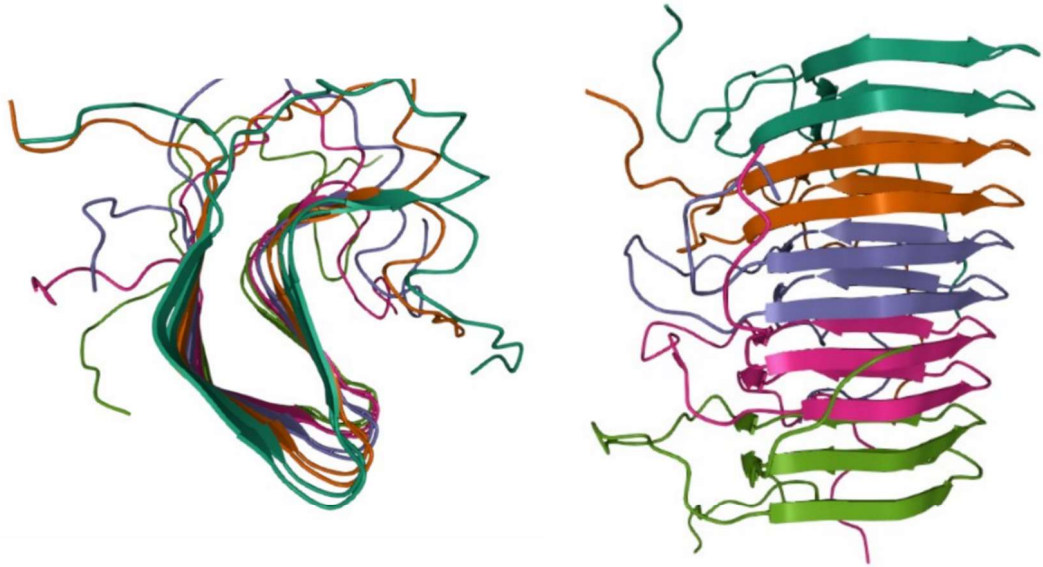

**C**

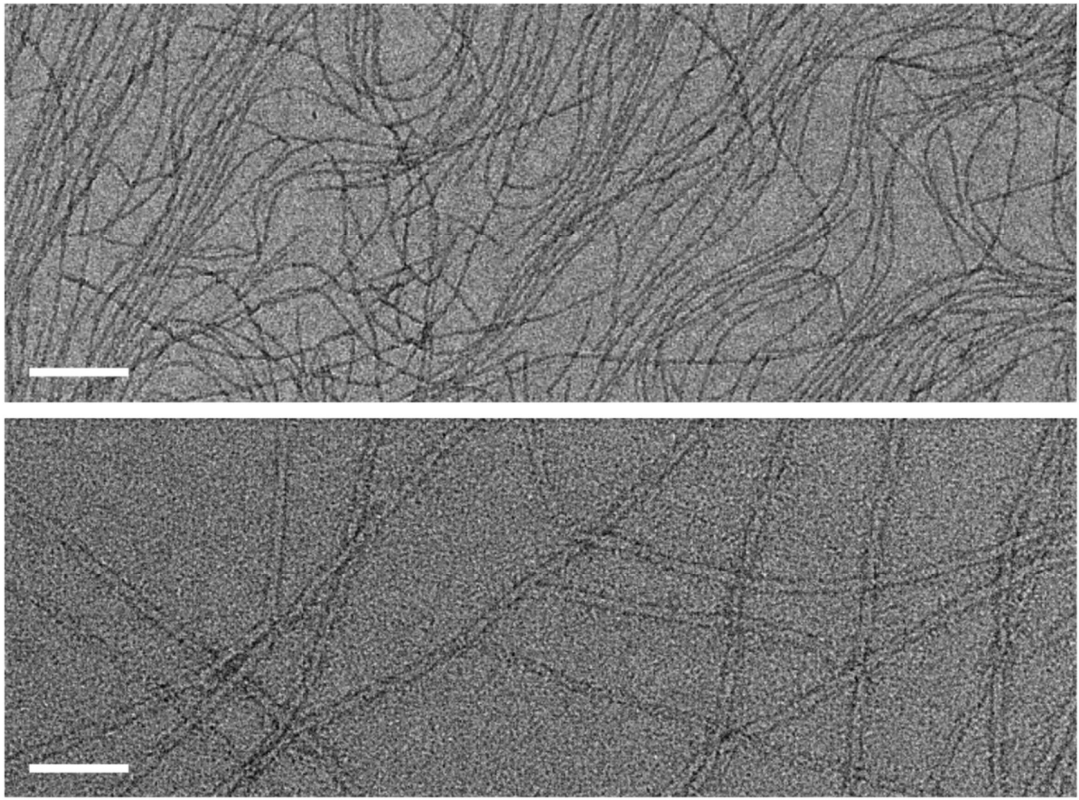

**Figure 1**

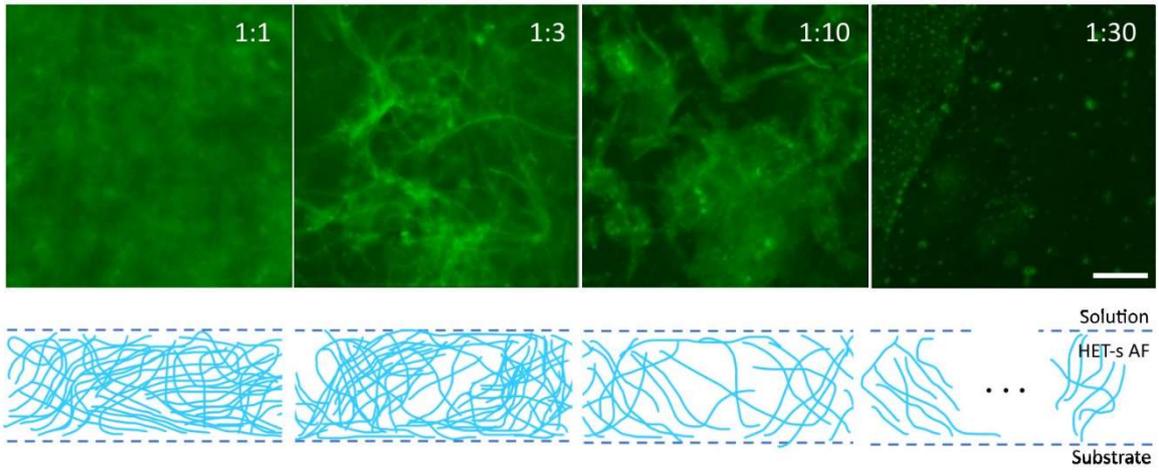

**Figure 2**

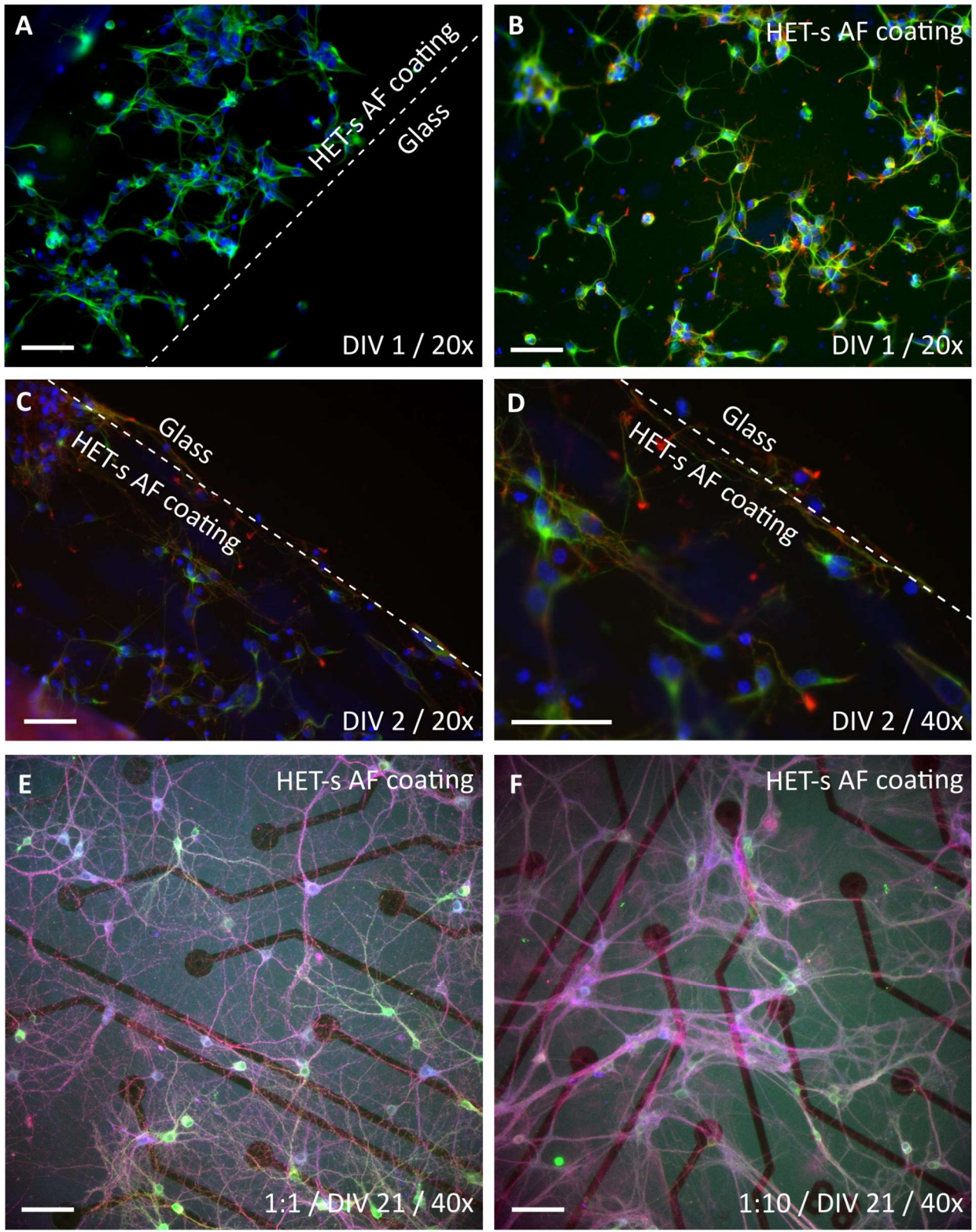

**Figure 3**

## A

| [HET-s] | 7 | 8 | 12 | 13 | 14 | 18 | 19 | 29 | >35 |
|---|---|---|---|---|---|---|---|---|---|
| 1:1 | 0 | 0 | 0 | 0 | 0 | - | - | - | - |
| 1:3 | 0 | 0 | 3 | 3 | 3 | - | - | - | - |
| 1:3 | 0 | 0 | 4 | 4 | 3 | 5 | 1 | - | - |
| 1:10 | 0 | 0 | 0 | 3 | 3 | 6 | 3 | - | - |
| 1:10 | 1 | 3 | 5 | 3 | 4 | 4 | 11 | 8 | 9 |
| 1:10 | 0 | 2 | 4 | 7 | 8 | 5 | 3 | 10 | - |
| 1:30 | 0 | 0 | 0 | 0 | 0 | - | - | - | - |

## B

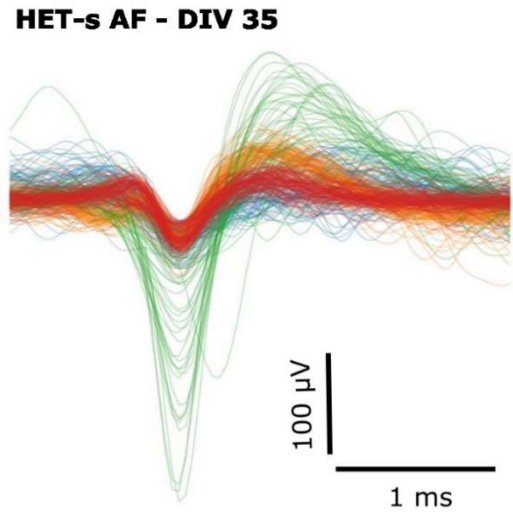

HET-s AF - DIV 35

## C

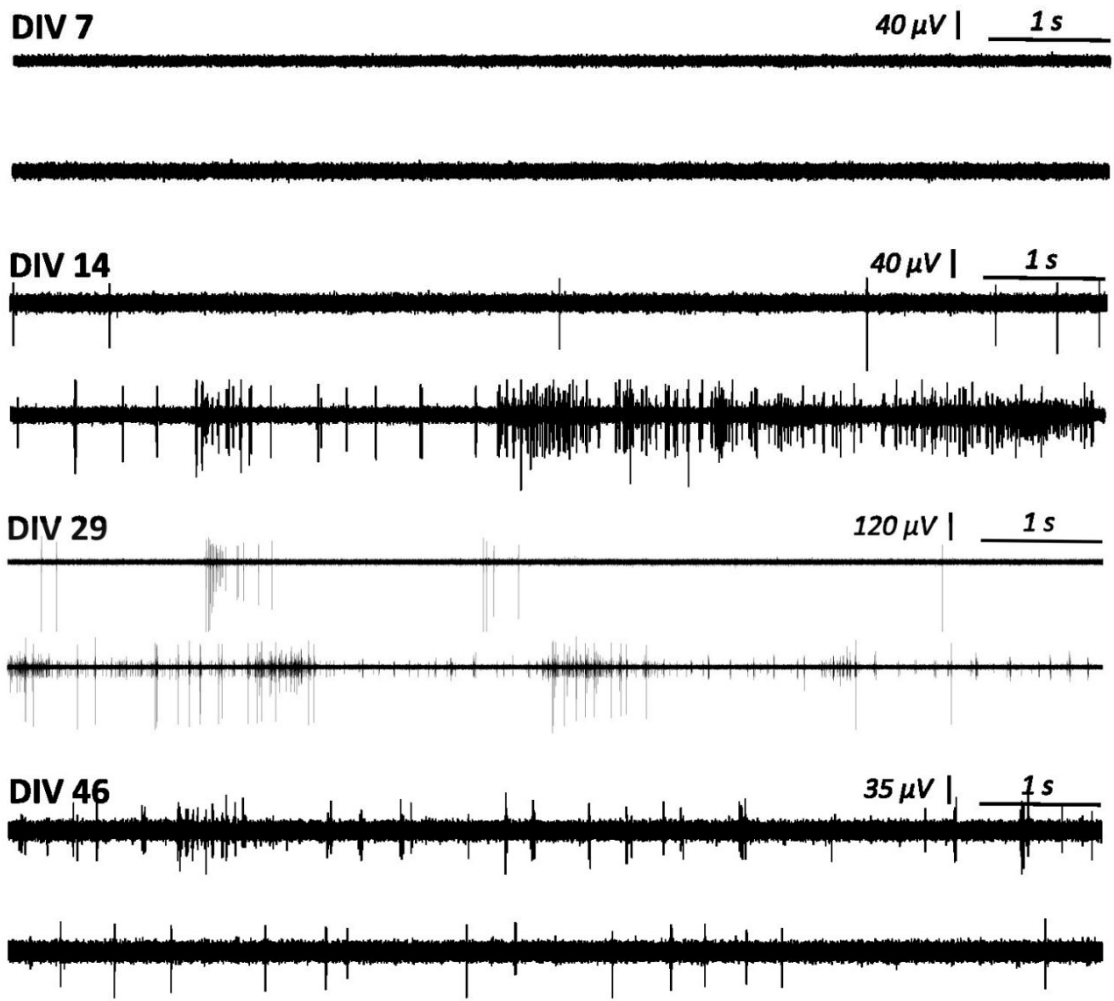

DIV 7   40 µV | 1 s

DIV 14   40 µV | 1 s

DIV 29   120 µV | 1 s

DIV 46   35 µV | 1 s

**Figure 4**

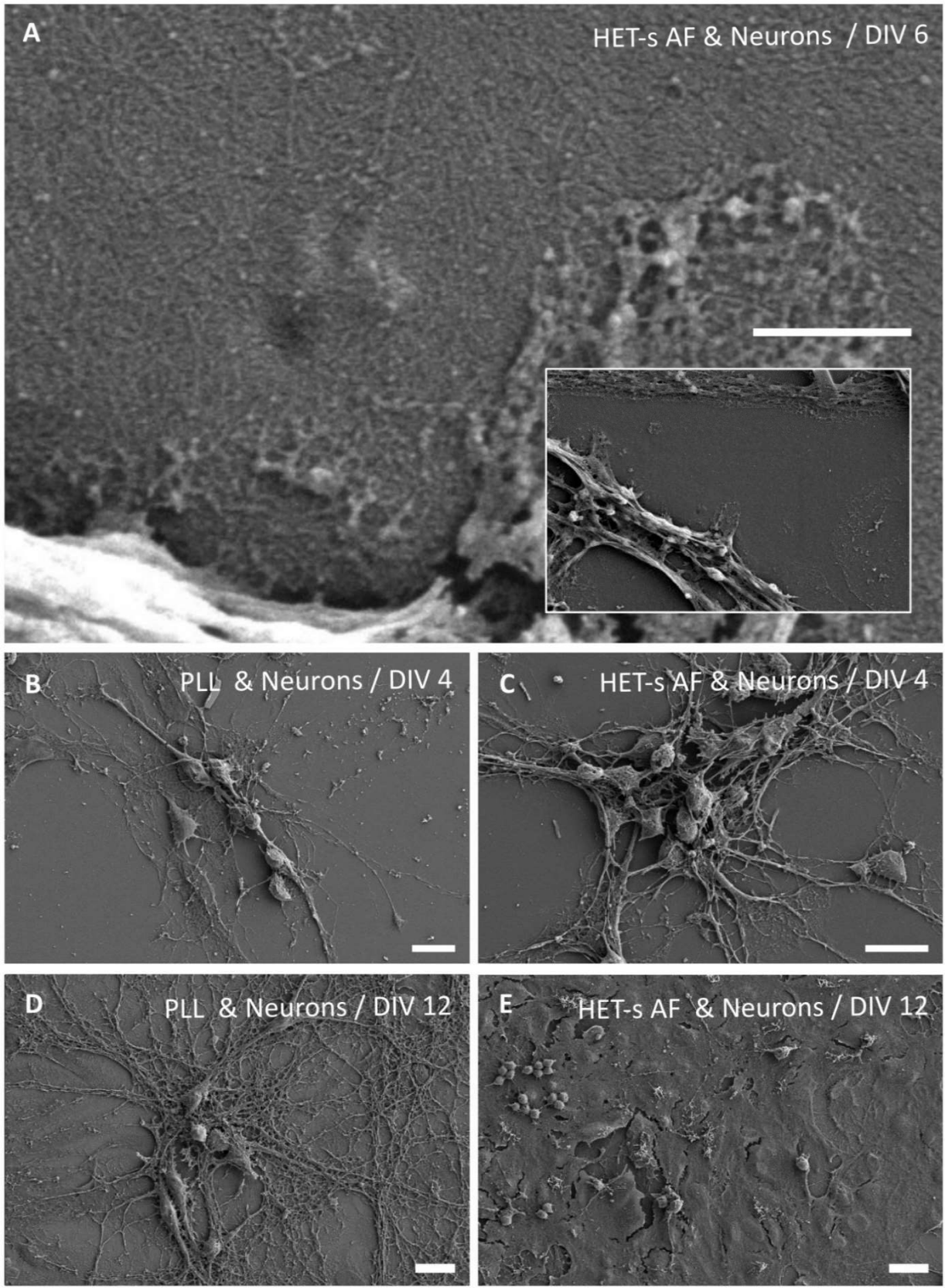

**Figure 5**

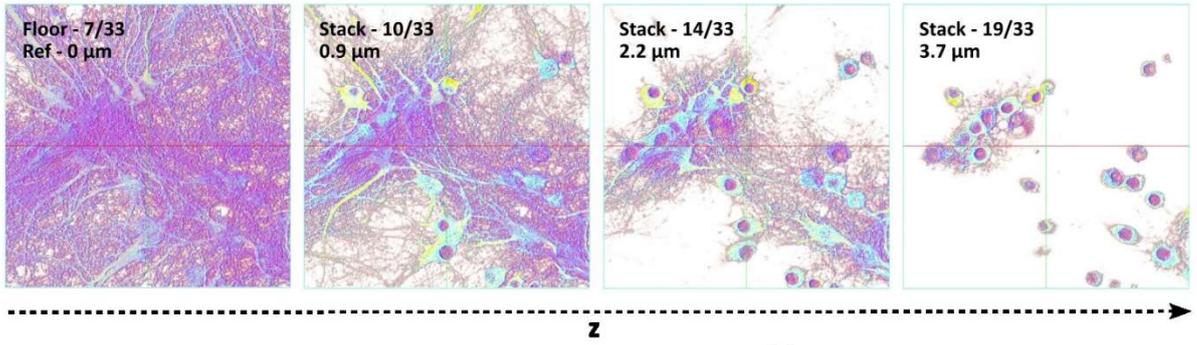
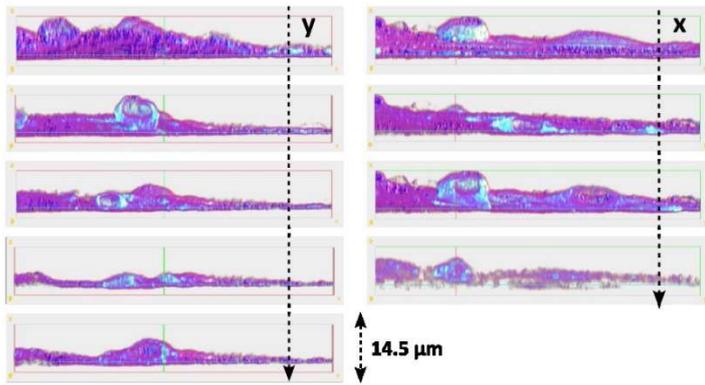
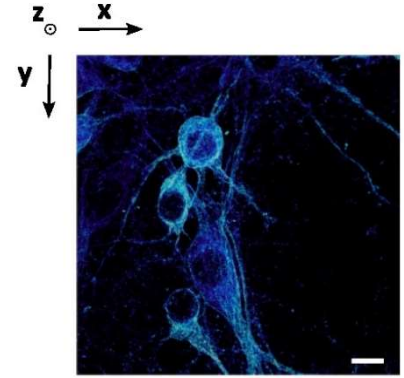

**Figure 6**

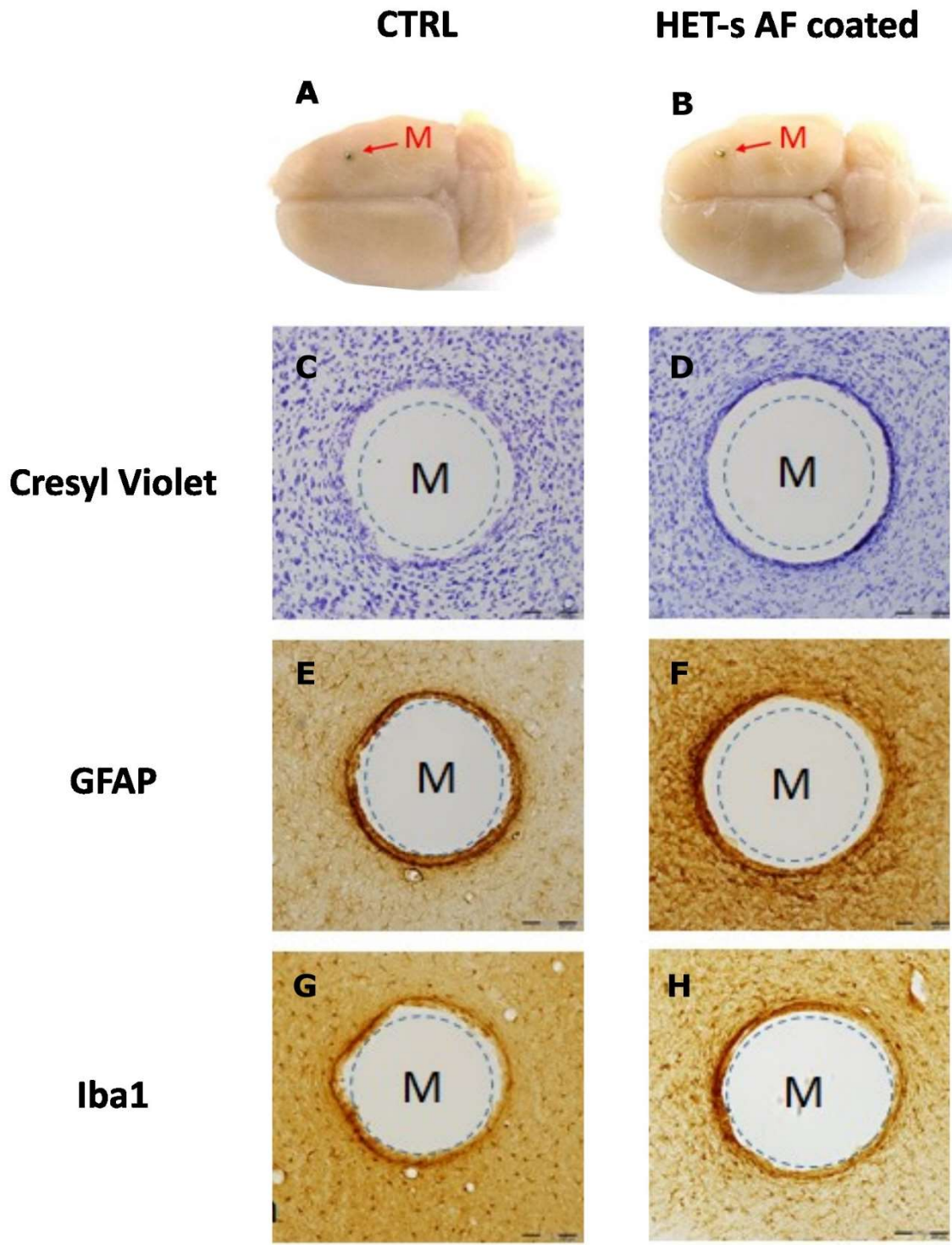

Figure 7